\documentclass[aps,pra,twocolumn,showpacs,preprintnumbers,amsmath,amssymb]{revtex4-2}
\usepackage{soul}
\usepackage{amsmath}
\usepackage{esvect}

\usepackage{natbib}
\usepackage{graphicx}
\usepackage{color}
\usepackage{tabularx}
\usepackage{multirow}

\usepackage{wasysym}

\usepackage{amssymb}

\usepackage[utf8]{inputenc}

\usepackage{float}

\usepackage{subfigure}

\usepackage{float}

\usepackage{hyperref}

\usepackage{cancel}

\usepackage{xcolor}

\usepackage{setspace}
\usepackage{soul}
\usepackage{wrapfig}

\definecolor{burntorange}{rgb}{0.8, 0.33, 0.0}

\begin{document}

\title{Single-frequency inverted Doppler-free resonance\\
as a platform for chip-scale optical clock}

\author{E.\,A.~Tsygankov$^{1}$}
\email[]{tsygankov.e.a@yandex.ru}
\author{D.\,S.~Chuchelov$^{1}$}
\author{M.\,I.~Vaskovskaya$^{1}$}
\author{V.\,V.~Vassiliev$^{1}$}
\author{S.\,A.~Zibrov$^{1}$}
\author{V.\,L.~Velichansky$^{1}$}
\affiliation{1. P.\,N. Lebedev Physical Institute of the Russian Academy of Sciences,\\
Leninsky Prospect 53, Moscow, 119991 Russia}

\newcommand{\red}{\textcolor{r}}
\newcommand{\blue}{\textcolor{blue}}

\begin{abstract}
We report on~the possibility to~obtain a~high-quality inverted Doppler-free resonance in~D$_1$ line of~alkali-metal atoms in~the single-frequency regime. The counter-propagating optical fields with linear and mutually orthogonal polarizations and the transition \hbox{$F_g=I+1/2\rightarrow F_e=I-1/2$} are proposed to~the use. We~establish the best possible nuclear spin value and the corresponding atomic isotope for this regime. Our experiment demonstrates that the resonance contrast-to-width ratio in~the single-frequency regime is~practically the same as~in~the dual-frequency regime, which simplifies the optical module to~be~used in~compact optical clocks. The achieved short-term frequency stability is~$3\cdot10^{-13}$ at~$1$~s, which is~comparable to~results that can be~obtained with the dual-frequency technique.
\end{abstract}

\maketitle

{\setstretch{0.99}Doppler-free spectroscopy~\cite{letokhov1977nonlinear,RevModPhys.54.697} of~the alkali-metal atoms D$_1$ line is~a~promising tool for the development of~emerging nowadays compact optical frequency standards. To~prevent the hyperfine optical pumping the counter-propagating bichromatic laser fields can be~used. In~this case, at~the exact optical resonance, mutually orthogonal polarizations ensure the absence of~non-absorbing superpositions of~states at~ground level~nS$_{1/2}$. On~the contrary, when the laser frequency is~detuned from that of~the optical transitions, the fields interact with different velocity groups of~atoms. They effectively induce the effect of~coherent population trapping (CPT) both at~microwave and Zeeman coherences, which leads to~a~significantly smaller level of~absorption compared to~the zero velocity group. As~a~result, there is~a~large dip at~the center of~a~suppressed Doppler contour. For the first time, this technique providing high-contrast Doppler-free absorption peaks in~caesium atoms was demonstrated in~\cite{Hafiz:16}.

The linewidth of~the inverted resonance could go~down to~about two-three natural linewidths of~the nP$_{1/2}$ state (where $n=5,6$ corresponds to~rubidium and caesium atoms, respectively) \hbox{($10-15$~MHz)}~\cite{brazhnikov2019dual,zhao2021laser,gusching2021short,gusching2023short}. The ratio of~the resonance width to~the frequency of~the transition of~the dual-frequency absorption resonance is~better than the reference lines of~microwave compact atomic clocks based on~the double radio-optical resonance and the CPT effect.
The frequency stability of~the systems providing feedback from the described above Doppler-free absorption peaks as~reference lines was measured in~\cite{Hafiz:16,brazhnikov2019dual,zhao2021laser,gusching2021short,gusching2023short}. The typical results for the measurement at~one second were in~the interval varied from $2\cdot10^{-12}$ to~$7\cdot10^{-12}$, while the best achieved result was $3\cdot10^{-13}$ at~$1$~s and $5\cdot10^{-14}$ at~$100$~s~\cite{gusching2023short}.

The drawback of~the dual-frequency scheme is~the requirement to~use an~external modulator or~a~diode laser with extended cavity. The latter provides a~high modulation efficiency,~i.e., a~low power of~the microwave~modulation is~required to~concentrate a~maximal possible amount of~optical power in~the sidebands used to~interrogate alkali-metal atoms. The modulator and cavity are bulky compared to~atomic cells with a~subcentimeter size, therefore limiting the clocks compactness. Another issue is~the presence of~microwave coherences. This leads to~a~periodical dependence of~the resonance amplitude on~the atomic cell position along the optical axis. The maximal signal is~achieved when the distance between the cell's center and a~retro-reflecting mirror is~equal to~the microwave transition wavelength. As~a~result, there is~a~trade-off between a~good signal: when the mirror is~placed right after the cell, a~device is~compact but the signal is~lower compared to~the case when the distance is~chosen to~provide its~maximal level.

In~this Letter, we~demonstrate that a~high-quality inverted Doppler-free resonance can be~obtained by~using a~monochromatic counter-propagating optical fields and the optical transition $F_g=I+1/2\rightarrow F_e=I-1/2$ of~an~alkali metal atom D$_1$ line, where $I$ is the nucleus spin. Experimentally, an~inverted resonance was observed at~the $F_g=2\rightarrow F_e=1$ transition in~$^{87}$Rb atoms. This possibility stems from the fact that there are only unperturbed Zeeman $\Lambda$-schemes providing the CPT effect for this transition and the branching of~the spontaneous decay to~ground-state hyperfine level $F_g=I-1/2$ is~smaller than $1/3$ for any $I$. In~the off-resonant case the counter-propagating fields induce the CPT effect for the two corresponding velocity groups. On~the contrary, there are no~Zeeman coherences at~the zero optical detuning if~polarizations of~the fields are linear and mutually orthogonal. Therefore, an~inverted Doppler-free resonance is~observed, whose amplitude does~not depend on~the position of~the atomic cell along the optical axis since microwave coherences are not involved. The described single-frequency scheme can be~implemented by~using a~diode laser without an~external cavity. The mirror and $\lambda/4$-plate can be~placed right after the atomic cell without the loss in~the signal. As~a~result, the physics package can be~compactified to~a~size comparable with that of~chip-scale microwave clocks based on~the CPT effect.}
\begin{figure}[t] 
  \center
  \includegraphics[width=0.5\textwidth]{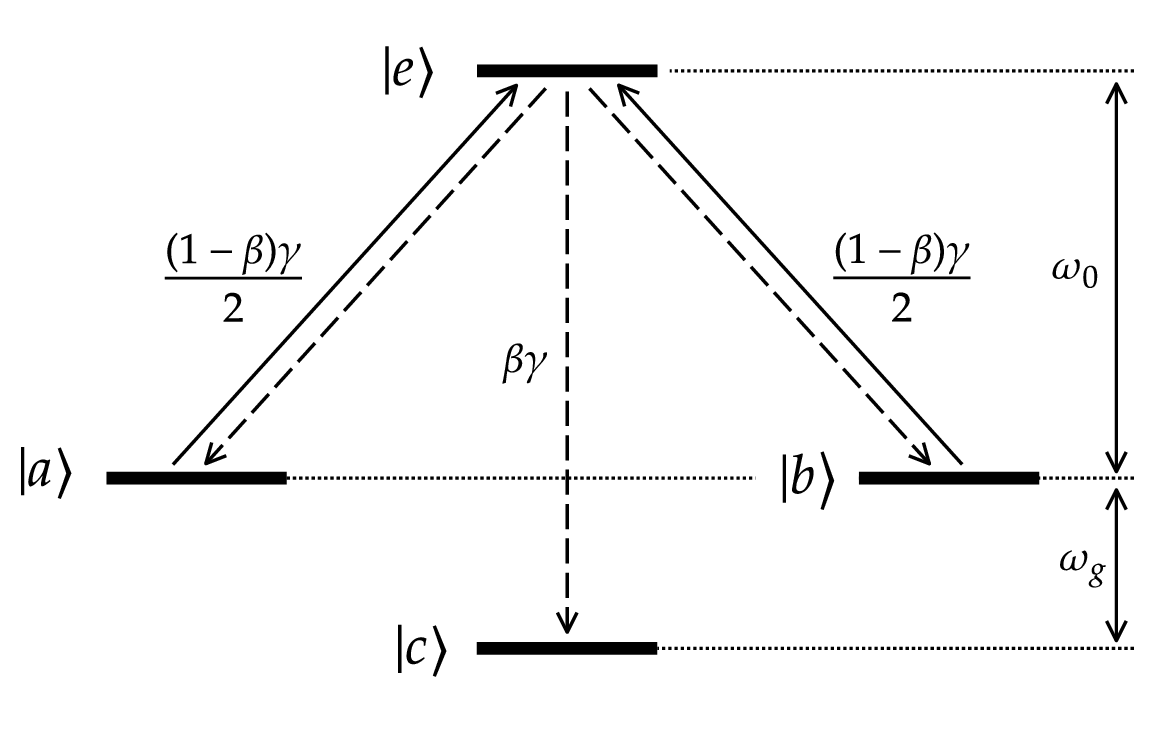}
  \caption{System of levels under consideration. $\gamma$ is the natural width of the excited state $|e\rangle$. The parameter $\beta$ determines branching of the system and amount of atoms decaying to non-absorbing level $|c\rangle$. The solid lines show the induced transitions and the dotted lines show the spontaneous decay.}
  \label{LevelsScheme}
\end{figure}

We~begin from an~investigation whether the D$_1$ line's transition \hbox{$F_g=I+1/2\rightarrow F_e=I-1/2$} can provide an~inverted Doppler-free resonance when there is~spontaneous decay to~a~nonabsorbing hyperfine level $F_g=I-1/2$. To~do~this, let us~consider the following pair of~counter-propagating monochromatic optical fields:
\begin{equation}
\mathcal{E}(t)=\mathcal{E}_0\left[\mathbf{e}_x\cos{\left(\omega_Lt-kv_z\right)}+\mathbf{e}_y\cos{\left(\omega_Lt+kv_z\right)}\right],
\end{equation}

\noindent where $\mathcal{E}_0$ is~assumed to~be~real, $\mathbf{e}_x$, $\mathbf{e}_y$ are the unity vectors and $\omega_L$ is~the field's frequency having detuning $\Delta_L$ from the spacing $\omega_0$ between levels $|a\rangle$, $|b\rangle$ and $|e\rangle$ of~the four-level scheme presented in Fig.~\ref{LevelsScheme}. $k$~is the wave vector and $v_z$ is~the atomic velocity along the optical axis $z$, which is~chosen as~the quantization axis. The component of~the optical field with right circular polarization couples levels $|a\rangle$ and $|e\rangle$, while the one with left circular polarization induces transitions between levels $|b\rangle$ and $|e\rangle$. The chosen system of~levels and transitions models the case \hbox{$F_g=I+1/2\rightarrow F_e=I-1/2$} since there is~the unperturbed $\Lambda$-scheme. This means that optical transitions from levels $|a\rangle$ and $|b\rangle$ occur only to~the common excited state $|e\rangle$. The level $|c\rangle$ is~nonabsorbing, since the ground-state splitting $\omega_g$ is~significantly greater than the Doppler width $\omega_D$ of~the optical transitions. This allows us~to model the hyperfine optical pumping of~the level $F_g=I-1/2$.

By~using the resonant (the rotating-wave) and small saturation approximations with the adiabatic elimination of~the excited state, the following set of~equations for the density matrix elements can be~obtained:
\begin{subequations}
\begin{equation}
\rho_{ee}=\mathcal{S}_+(g+2\rho_{ab})+\mathcal{S}_-(g-2\rho_{ab}),
\end{equation}
\begin{equation}
\dfrac{\partial}{\partial t}g=-\beta\gamma\left[\mathcal{S}_+\left(g+2\rho_{ab}\right)+\mathcal{S}_-\left(g-2\rho_{ab}\right)\right],
\end{equation}
\begin{equation}
\dfrac{\partial}{\partial t}\rho_{ab}=-\gamma\left(\mathcal{S}_++\mathcal{S}_-\right)\rho_{ab}-\gamma\left(\mathcal{S}_+-\mathcal{S}_-\right)g/2,
\label{coherenceEq}
\end{equation}
\label{Eqs}
\end{subequations}

\noindent with the conservation law $\rho_{cc}=1-g$. In~the equations above \hbox{$\mathcal{S}_{\pm}=V^2/\left[\gamma^2/4+\left(\Delta_L\pm kv_z\right)^2\right]$}, where \hbox{$V=d\mathcal{E}_0/\sqrt{2}\hbar$} is~the Rabi frequency, \hbox{$\Delta_L=\omega_L-\omega_0$}. For convenience, the equations are written for $g$, the sum of~levels $|a\rangle$ and $|b\rangle$ populations: $g=\rho_{aa}+\rho_{bb}$. Since the optical field induces the exact two-photon resonance, the imaginary part of~the non-diagonal element $\rho_{ab}$ is~absent and an~equation for $\rho_{ba}$ is~not required. As~one can see, there~is the nonhomogeneous term \hbox{$\propto\left(\mathcal{S}_+-\mathcal{S}_-\right)$}~in the equation for $\rho_{ab}$. The opposite signs for $\mathcal{S}_{\pm}$ in~the last term of~\eqref{coherenceEq} stem from the fact that the fields polarizations are orthogonal. The circular components acquire the phase factors $\pm\pi/2$ resulting in~their difference equal to~$\pi$ and the negative sign for $\mathcal{S}_-$.
\begin{figure}[t] 
  \centering
  \includegraphics[width=0.5\textwidth]{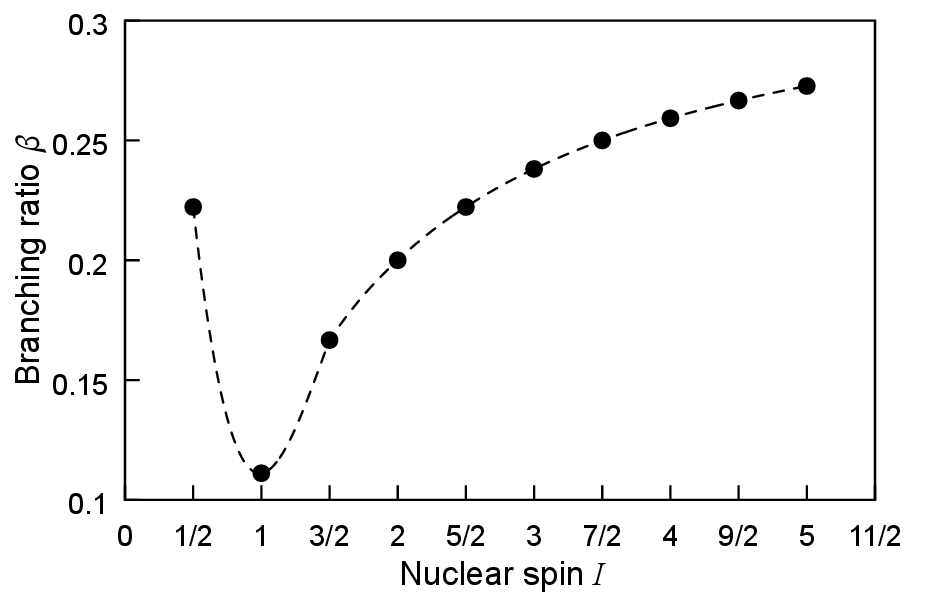}
  \caption{Dots: dependence of the branching coefficient on the nuclear spin for the spontaneous decay of the alkali-metal atom D$_1$ line excited-state level $F_e=I-1/2$ to the ground-state level $F_g=I-1/2$. The dashed line is the guide for eyes.}
  \label{BranchingRatioFig}
\end{figure}

Firstly, we~investigate the point $\Delta_L=0$, where \hbox{$\mathcal{S}_+-\mathcal{S}_-=0$}, i.e., where the CPT effect is~absent. In~this case the averaged excited-state population over time interval $\tau$, which we~treat as~the mean flight duration of~the atom through the optical fields beam, is~proportional to~\hbox{$\beta^{-1}\left[1-\exp{\left(-\beta P\tau\right)}\right]\equiv\beta^{-1}\left\{1-\exp{\left[-\beta(1-\beta)P_{r}\tau/2\right]}\right\}$}, where $P=V^2/\gamma$ and \hbox{$P_{r}=V^2_{r}/\gamma$}. Here the Rabi frequency $V_{r}$ contains the reduced dipole matrix element, \hbox{$d=(1-\beta)d_{r}/2$}. Such a~replacement is~valid since the spontaneous decay channels are \hbox{$(1-\beta)\gamma/2$} and \hbox{$\gamma\propto d^2_{r}$}. As~can be~seen, the absorption coefficient determined by~the excited-state population is~maximal at~$\beta=0$, which is~an~expected result.

Secondly, we~consider the points where the condition \hbox{$\gamma\ll|\Delta|_L\ll\omega_D$} holds. Thus, the optical fields interact with different velocity groups of~atoms and change of~their number due to~the distribution on~the longitudinal velocity can be~neglected compared to~the first case. The averaged over $\tau$ excited-state population now is~\hbox{$\propto(1+\beta)^{-1}\left\{1-\exp{\left[-\left(1-\beta^2\right)P_{r}\tau/2\right]}\right\}$}. This expression is~the smallest at~$\beta=1$, which means that the absorption is~smaller at~more forbidden transitions than at~open but with presence of~dark coherence $\rho_{ab}$.

Thus, the maximal absorption for $\Delta_L=0$ and the minimal absorption in~the non-resonant case are observed at~$\beta=0$ and $\beta\neq0$, respectively. However, the comparison shows that the inverted Doppler-free resonance has the greatest amplitude at~$\beta=0$. Therefore, it~will be~more sharp at~the lower value of~$\beta$, which also is~expected result. What is~surprising is~that the resonance will be~inverted even at~small values of~$\beta$ close to~$0$.

In~the following step, we~investigate how the value of~$\beta$ depends on~the nuclear spin $I$. Due to~the symmetry of~the dipole operator, each magnetic sublevel of~the excited-state level \hbox{$F_e=I-1/2$} spontaneously decays to~the ground-state level \hbox{$F_g=I-1/2$} with the same branching coefficient $\beta$ which is~given by
\begin{equation}
\beta=\left(2F_g+1\right)\left(2J_e+1\right)\left(\begin{Bmatrix}
J_g & J_e & 1 \\
F_e & F_g & I
\end{Bmatrix}\right)^2,
\end{equation}

\noindent where the braces denote the Wigner's $6$-j symbol. We~consider the case \hbox{$J_e=J_g=1/2$}, \hbox{$F_e=F_g=I-1/2$} and get
\begin{equation}
\beta=\dfrac13\cdot\dfrac{2I-1}{2I+1},
\end{equation}

\noindent where $I\ge1/2$.

The dependence of~$\beta$ on~$I$ is~given in~Fig.~\ref{BranchingRatioFig} for convenience. Despite $\beta=0$ for $I=1/2$, we~have replaced this value by~$2/9$ for the following reason. In~this case the decay \hbox{$F_e=0,\,m_{F_e}=0\rightarrow F_g=0,\,m_{F_g}=0$} is~forbidden in~the electric-dipole approximation, but the magnetic sublevel \hbox{$F_g=1,\,m_{F_g}=0$} is~nonabsorbing for the optical fields due to~the selection rules. The corresponding spontaneous decay rate of~the excited state $F_e=0$ to~this sublevel is~$2/9$ of~the natural width.

The minimal value of~$\beta=1/9$ is~for fermion atoms with \hbox{$I=1$} [formally, for $S=1$ (atoms with two electrons in~the outer shell) and $I=1/2$ the momentum $F$ is~also $1/2$ and $3/2$, but coefficient $\beta$ is~greater and equal to~$1/3$]. However, the corresponding stable isotope here is~only $^6$Li. Its~hyperfine splitting of~the state $2$P$_{1/2}$ is~significantly smaller than the Doppler width and the transition \hbox{$F_g=3/2\rightarrow F_e=1/2$} is~not resolved. Therefore, atoms of~$^6$ Li are not suitable to~obtain an~inverted Doppler-free resonance in~the single-frequency regime and also unsuitable due to~its chemical properties. Then we~have $\beta=1/6$ for $I=3/2$, which corresponds to~$^{87}$Rb, $^{39,\,41}$K. The hyperfine splitting here is~greater than $\omega_D$ only for $^{87}$Rb, which turns out to~be the most suitable atom for inverted Doppler-free resonance in~the single-frequency regime. The next atoms are $^{85}$Rb ($I=5/2$) and $^{133}$Cs ($I=7/2$). But the $^{85}$Rb does not have a~well-resolved hyperfine structure of~the state $5$P$_{1/2}$ while the value of $\beta$ is $1.5$ times greater for $^{133}$Cs latter compared to~$^{87}$Rb.
\begin{figure}[b] 
  \center
  \includegraphics[width=0.5\textwidth]{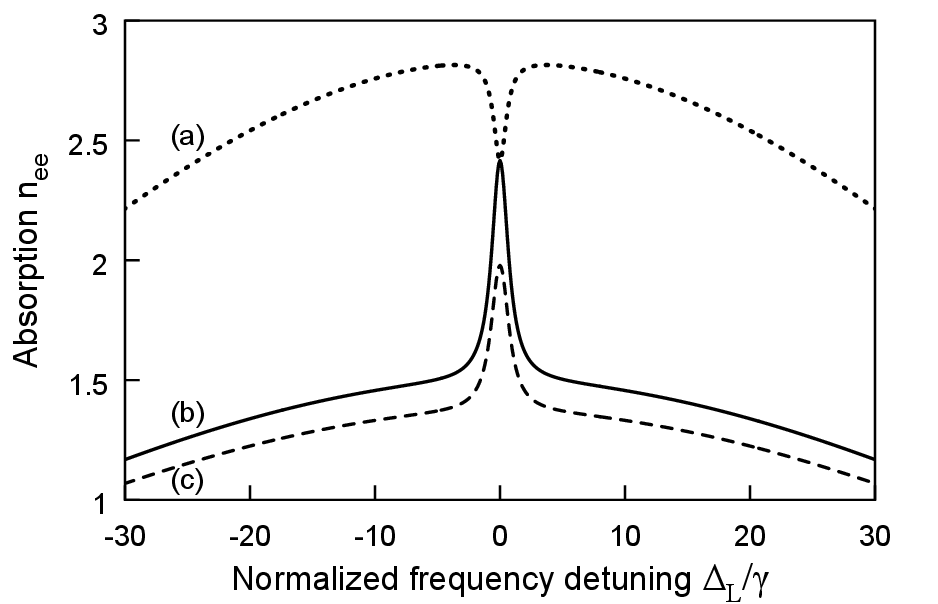}
  \caption{Doppler-free resonance for $\beta=1/6$ (two top curves) and $\beta=1/4$ (bottom curve). The CPT effect is~neglected for the upper curve and accounted for two others. The vertical axis is~given in~units $(V_{r}/\gamma)^2\cdot\gamma/\sqrt{\pi}kv_p$. Optical pumping factor $P_{r}\tau/2\pi$ is~equal to~$1/2$, $kv_p/\gamma=60$, which corresponds to~$^{87}$Rb atoms at~the temperature of~$300$~K.}
  \label{Single-FrequencyResonanceFig}
\end{figure}

To~demonstrate the principle, three curves of~Doppler-free resonance are given in~Fig.~\ref{Single-FrequencyResonanceFig}. They were plotted by~obtaining solution of~system~\eqref{Eqs} and averaging
\hbox{$n_{ee}\equiv\sqrt{1/\pi v^2_p}\exp{\left[-\left(v_z/v_p\right)^2\right]}\rho_{ee}(v_z,\,t)$} over the flight time $\tau$ with numerical integration over $kv_z/\gamma$ in~the interval $\left[-150,\,150\right]$. Here $v_p$ is~the most probable velocity. As~one can see, the CPT effect provides inverted Doppler-free resonance despite that there is~nonabsorbing sublevel $|c\rangle$ in~the system.

\begin{figure}[t] 
  \includegraphics[width=1\linewidth]{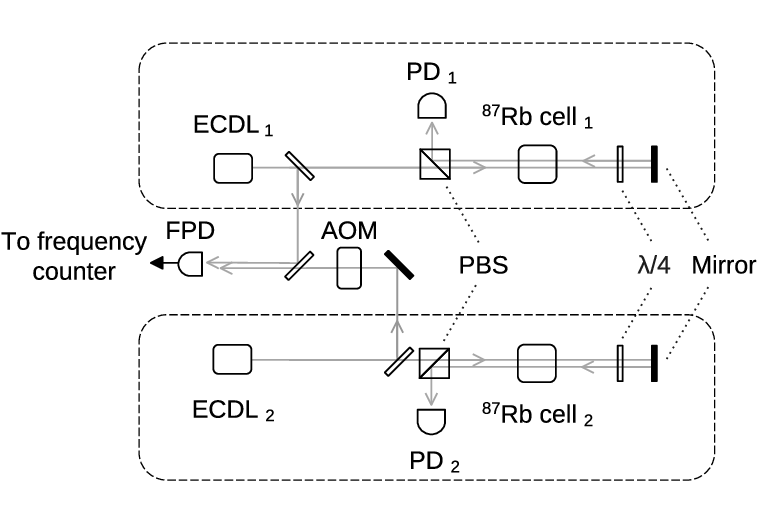}
  \caption{The experimental setup. ECDL---extended-cavity diode laser, PBS---polarizing beam splitter, \hbox{PD---photodetector}, FPD---fast photodetector, \hbox{$\lambda/4$---quarter-wave plate.}}
  \label{ExpSetup}
\end{figure}

We~experimentally investigated the inverted Doppler-free resonance in~$^{87}$Rb using a~single-frequency laser field. 
Fig.~\ref{ExpSetup} shows the schematic of~our experimental setup. 
Two identical systems with extended cavity diode lasers were implemented. They included interference filter for coarse radiation frequency tuning and cat’s eye as~an~output mirror~\cite{vassiliev2019vibration}.
The mirror was mounted on~piezoelectric transducer which provides fine frequency tuning. 
The laser radiation passed through an~optical isolator and a~combination of~a~half-wave plate and a~polarizing beam splitting cube. 
Then it~was directed to~the atomic cell, reflected back by~a~mirror, and registered by~a~photodetector.  
The linear polarizations of~the direct and reflected beams were orthogonal due to~a~quarter-wave plate placed in~front of~the mirror. 
The cylindrical atomic cell ($20$~mm in~diameter, $10$~mm in~length) filled with isotopically enriched $^{87}$Rb was placed inside a~three-layer magnetic shield to~suppress the influence of~the external magnetic field. 
The cell temperature was maintained close to~$60$~$^\circ$C with an~accuracy of~$0.01$~$^\circ$C. 

\begin{figure}[t] 
  \includegraphics[width=0.5\textwidth]{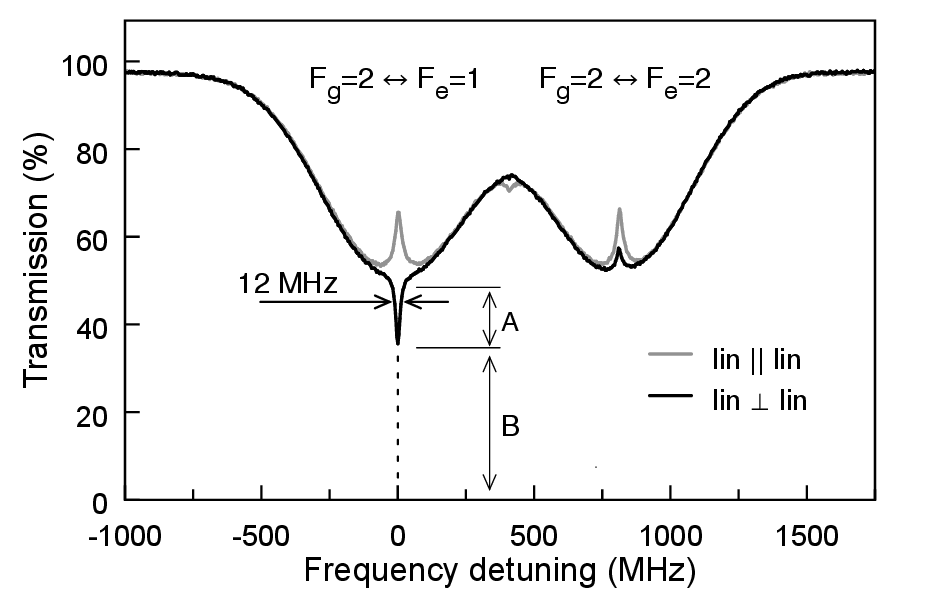}
  \caption{Doppler-free spectra of~$^{87}$Rb D$_1$ line taken at~crossed polarizations of~the counter-propagating waves. The horizontal axis represents the laser frequency detuning from the \hbox{$F_g=2\rightarrow F_e=1$} transition.}
  \label{Resonance}
\end{figure}

The Doppler-free resonance obtained at~a~laser intensity of~$\simeq 0.8$~mW/cm$^2$ are~shown in~Fig.~\ref{Resonance}. The $F_g=2 \rightarrow F_e=1$  absorption peak has width of~$12$~MHz and contrast ($C=A/B$) of~$30\%$.

Frequency stabilization to~the described resonance was performed in~a~usual way.
To~generate the error signal the laser current was modulated at~frequencies of~$60$ and $70$~kHz.
After synchronous detection, the demodulated transmission signal was processed by~the proportional–integral–derivative controller to~produce a~signal, which was applied to~the control inputs of~both the piezoelectric transducer and the laser current.
To~measure the frequency stability of~a~laser locked to~Doppler-free resonance the beatnote frequency of~two systems were registered.
A~part of~the radiation was split off from each laser system and sent to~a~fast photodetector.
To~convert the beat frequency from near-zero to~the high-frequency region, an~acousto-optic modulator was included in~one of~the systems, shifting the laser frequency by~$80$~MHz.
The beatnote frequency was registered with a~frequency counter.
\begin{figure}[b] 
  \includegraphics[width=0.5\textwidth]{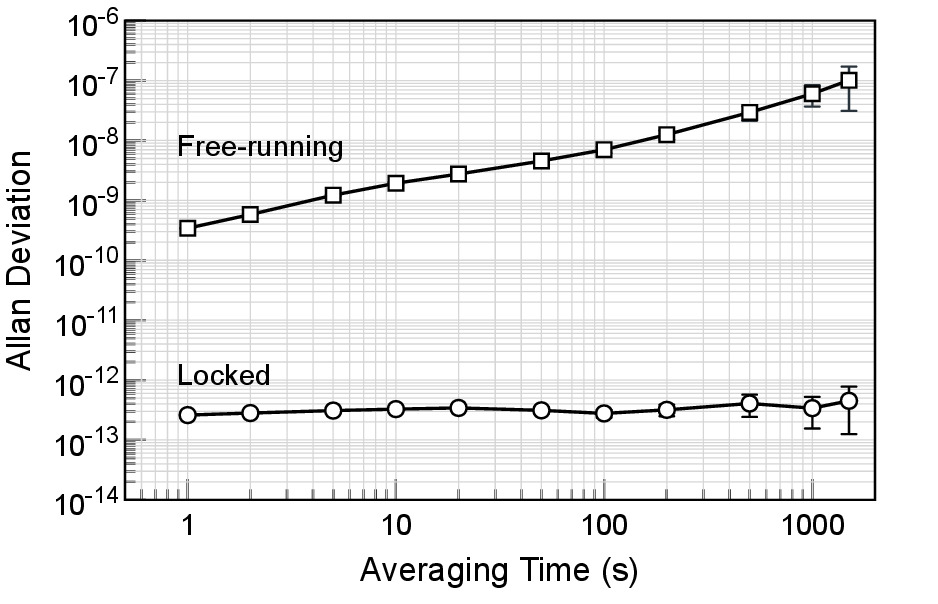}
  \caption{Allan deviation of~the laser beatnote frequency when one laser is~free-running (squares) and both lasers are locked (circles).}
  \label{stability}
\end{figure}

Two experiments were conducted. In~one of~them both lasers were locked to~$F_g=2 \rightarrow F_e=1$ resonance. In~the second experiment one of~the lasers was free running. Allan deviation of~the laser beatnote frequency for two regimes is~shown in~Fig.~\ref{stability}.
In~free-running mode, the frequency stability of~the laser beatnote is~$3.4\cdot10^{-10}$ at~$1$~second, which degrades to~$6\cdot10^{-8}$ at~$10^3$ seconds, primarily due to~temperature fluctuations and variations in~the laser resonator length.
In~locked mode, a~frequency stability of~$3\cdot10^{-13}$ at~$1$~second was achieved, which was limited by~the laser's frequency noise at~the modulation frequency. 
The short-term frequency stability can be~further improved by~optimizing experimental parameters, including cell temperature, radiation intensity, modulation frequency, and the feedback bandwidth.

\textbf{To~conclude.} We~proposed a~scheme to~obtain an~inverted Doppler-free resonance using a~monochromatic counter-propagating optical waves. The $^{87}$ atoms and transition $F_g=2\rightarrow F_e=1$ were established as~the best choice for this case. We~experimentally obtained the Doppler-free resonance with the similar quality that can be~obtained using a~bichromatic radiation. The attained frequency stability and the possibility to~radically decrease dimensions of~the facility (no~need for the Mach-Zehnder modulator) makes it~attractive for a~compact optical frequency standard.

\subsection*{Acknowledgments}

The authors receive funding from Russian Science Foundation (grant No. 24-72-10134).

\bibliographystyle{apsrev4-1}
\bibliography{references}

\end{document}